\documentstyle[11pt,epsfig]{article}
\newcommand{\be}{\begin{equation}}
\newcommand{\ee}{\end{equation}}

\newcommand{\beq}{\begin{equation}}
\newcommand{\eeq}{\end{equation}}

\newcommand{\bea}{\begin{eqnarray}}
\newcommand{\eea}{\end{eqnarray}}

\newcommand{\beqn}{\begin{equation}}

\newcommand{\eeqn}{\end{equation}}

\begin{document}
DESY 03-048
\hfill ISSN 0418--9833 \\

\vspace*{0.5 cm}
\begin{center}
\begin{Large}
\vspace*{0.5cm}
{\bf
The Process $\gamma(^*)+p \to \eta_c +X$:\\
A Test for the Perturbative QCD Odderon}
\\
\end{Large}

\vspace{0.5cm}
J. Bartels$^a$,
M.A.Braun$^b$
and G.P.Vacca$^c$ \\
$^a$ II. Inst. f. Theoretische Physik,
Univ.
Hamburg, Luruper Chaussee 149, D-22761 Hamburg\\
$^b$ St.Petersburg
University,
     Petrodvoretz, Ulyanovskaya 1, 198504, Russia\\
$^c$ Dipartimento di Fisica - Universit\`a di Bologna and
INFN -  Sezione di Bologna,\\
via Irnerio 46, 40126 Bologna, Italy
\end{center}
\vspace*{0.75cm}


\begin{abstract}
\noindent
The  rates of inclusive photo- and electroproduction of the $\eta_c$
meson: $\gamma(^*)+p \to \eta_c + X$ are calculated in the triple Regge
region, integrated over the diffractive mass $X$.
For the Regge exchanges we use
the hard pomeron and odderon, both being calculated in the framework of
perturbative QCD. The integrated cross-section depends upon
the coupling of the BFKL pomeron to two $C=-1$ odderons, and it is found
to be of the order 60 pb for photoproduction and 1.5 pb at
$Q^2=25$ GeV$^2$.

\end{abstract}


\section{Introduction}
The existence of the odderon ~\cite{Lukaszuk:1973nt},
the partner of the Pomeron which is odd under charge conjugation $C$, is an
important prediction of perturbative QCD.
It is a direct consequence of the number of colours $N_c$ being greater than
two. In the leading order, the odderon
appears as a bound state of three reggeized gluons.
Its experimental observation is a strong challenge for the experimentalists.
A particular promising scattering process where the exchange of the odderon
may be seen is the diffractive production of particles with a C-odd
exchange, such as photo- and electroproduction of pseudoscalar mesons (PS),
provided a large momentum scale is involved, which
gives a justification for the use of perturbative QCD.
This includes, in particular, the diffractive production of charmed
pseudoscalar mesons, for example the process $\gamma + p \to \eta_c+ p$.
Correspondingly, a large amount of literature has been devoted
to this class of diffractive processes. For large photon
virtualities $Q^2$, for heavy mass PS mesons (such as $\eta_c$), and for
large momentum transfers the relevant impact factors for the
transition $\gamma (\gamma^*) \to$PS have been be calculated perturbatively
~\cite{Czyzewski:1997bv}.
As to the odderon structure, different models have been used:
the exchange of three noninteracting gluons in a \mbox{C=-1} state
~\cite{Czyzewski:1997bv}, and, more recently,
the perturbative QCD odderon with intercept exactly unity
~\cite{Bartels:2000yt, Vacca:2000bk}, used to calculate the production
rates of the $\eta_c$ in \cite{BBCV}
\footnote {A different approach, a
nonperturbative odderon based  upon the idea of a ``stochastic QCD vacuum''
has been used in  ~\cite{Heidelberg}.}.
In both models, there is some uncertainty coming from the coupling of the
odderon to the proton. Numerical estimates for the cross sections
turn out to be somewhat different in these approaches.
However in all cases they are very small and, most unfortunately,
do not grow with energy (in the case of the perturbative QCD
odderon, they even slowly decrease with energy). This leaves little hope to
see the odderon by raising the energy of the reaction.

However the situation may become different if, instead of the quasielastic
process $\gamma + p \to \eta_c + p$, one considers the inclusive cross
section $\gamma + p \to \eta_c + X$ in the triple Regge region.
In this case the odderon does not couple directly
to the quarks of the target proton but rather to the diffractive system
$'X'$ which, for high masses, can modelled by a cut gluon ladder, the
gluon density inside the proton. The proton is therefore coupled to the
cut gluon ladder, i.e. the Pomeron, and this coupling is known
through the gluon density. This fact permits to avoid
the previously mentioned uncertainties in the odderon-proton coupling.
Together with this process also the low mass diffractive state (the proton)
with the meaning of a double odderon exchange is usually considered.

In the Regge language this new situation basically involves
the coupling of two odderons to a cut Pomeron, the POO vertex.
Since we are using perturbative QCD both for the odderon and for the cut
Pomeron, also this vertex has to be calculated in perturbative QCD.
This has been done in ~\cite{BE}: the vertex has been obtained in
an analysis of a six gluon amplitude $D_6$. In our application of this vertex
we shall restrict ourselves to the leading
large-$N_c$ limit, which leads to a relatively simple form of $D_6$.
In ~\cite{BE} it was also shown that the full amplitude $D_6$ can be
decomposed into the sum of two contributions, where the first one
results from the reggeization of the gluon and , and the second one
contains the POO vertex. Correspondingly, also our cross section
comes as the sum of two pieces (denoted by $P$ and $POO$, respectively).
The second one corresponds to the normal  `triple Regge picture' where
the Pomeron splits into two odderons, whereas the first one is related
to reggeization of the gluon and leads the exchange of three
noninteraction gluons in the odderon channel.
We will calculate both of them.
For the odderon states we will use the solution found in
~\cite{Bartels:2000yt}, which has a maximal intercept (unity) and a very
simple analytical form. This solution has already been used by us to
calculate the odderon exchange in the  process
$\gamma + p \to \eta_c + p$ \cite{BBCV}.

The use of these elements allows to compute the diffractive cross section
$\frac{d^2 \sigma}{dt dM^2}$ for the process $\gamma + p \to \eta_c + X$.
Our perturbative QCD analysis only depends upon one free parameter, the
coupling of the gluon ladder to the proton: this coupling will be fixed
by fitting the model to the gluon density of the proton. To simplify
the calculations we restrict ourselves to the integrated (over $M^2$)
cross section: the calculation of the differential (in $M^2$) cross
section requires a slightly different treatment of the $D_6$ amplitude
which will not be pursued in the present work.
Nevertheless the most inportant and basic information is given by the
integrated cross section.

As we have indicated before, we expect that the cross section for the
inclusive process $\gamma + p \to \eta_c + X$ is larger than that for the
quasielastic process  $\gamma + p \to \eta_c + p$.
We know that the cut gluon ladder grows as $\exp \Delta y$, where $y$ is the
rapidity gap between the proton and the POO vertex, and  $\Delta$ is the
value of the Pomeron intercept minus unity. From this it follows that the
bulk of the inclusive cross-section will come from the region where $y$ is as
large as possible. i.e. close to the total rapidity of the process
(note however, that, in order to see the exchange of an odderon,
one needs also a large rapidity gap between the outgoing $\eta_c$ and the
diffractive system). In other words, the mass of the diffractive system ``X''
wants to become as large as possible.
Because of this growth (with energy) of the cut gluon ladder
we expect to see a strong enhancement of the
inclusive cross-section at high energies, compared to the quasielastic
process $\gamma + p \to \eta_c + p$ where the gluon ladder is absent.
The comparison of our results with the quasielastic cross section
is made difficult by the intrinsic uncertainty of the odderon-proton
coupling: a recent analysis shows that the estimate obtained in
~\cite{Czyzewski:1997bv} and also adopted in
~\cite{BBCV} may have used a too large value of this coupling and has to be
reduced: if this is the case, the cross section obtained in the present paper
is, in fact, much larger that the quasielastic one.

Our paper is organized as follows.
In the next section we shall briefly recall some results which constitute
our starting point to attack the problem, the cross section formula.
In section 3 the first contribution (P) is considered, and the
corresponding integrated (in $M^2$) cross section is written in terms of a
multidimensional integral which, later on, will be be computed numerically.
In sections 4 and 5 the structure of the second contribution (POO) is
considered. Both contributions are calculated in the large $N_c$ limit, which
leads to considerable simplifications and shows a symmetry
shared by the leading odderon states \cite{Bartels:2000yt}.
Finally, the numerical analysis is presented and discussed in section 6,
followed by the conclusions.

\section{The cross-section formula in perturbative QCD}.

We start from the analysis \cite{BE} of QCD Feynman
diagrams in the leading $\log s$ approximation, and we recapitulate
the main results. The approach taken from ~\cite{BE} is
a generalization of a previous analysis~\cite{BW} of the $4$ gluon system
(related to the triple pomeron vertex): it extends this analysis
up to $6$ gluons in the $t$ channel, and so it encounters, for the first time,
the two-odderon state. As described in ~\cite{BE}, the analysis of Feynman
diagrams in the high energy limit leads to a tower of gluon
amplitudes, $D_2$, $D_3$, $D_4$, $D_5$, and $D_6$, which satisfy a set of
coupled integral  equations. These functions are non-amputated, i.e.
they contain reggeon denominators for the outgoing (reggeized)
gluon states (see Ref.~\cite{BW}). The latter
are more convenient degrees of freedom than the elementary gluons in this
kinematics. In the present context we are interested in the $D_6$ amplitude
which can be used to build the cross section, integrated in the diffractive
mass.

In the analysis in refs.~\cite{BE} and~\cite{BW},
all functions $D_2,...$ start from
the impact factor of a virtual photon which splits into a
quark-antiquark pair. In the present case, the external particle
is the proton: assuming that, to a good approximation, the proton
can be viewed as a quark-diquark system, the coupling of the gluons
to the proton should have the same structure
as in the photon case; only the overall normalization of this coupling
has to be treated as a phenomenological parameter.

The diagrammatic structure of the differential cross section
$\frac{d^2 \sigma}{dt dM^2}$ for the process $\gamma + p \to \eta_c + X$
is illustrated in Fig.1a. In the two exchange channels one recognizes
the two odderon states, consisting of three gluons with pairwise
interactions. The structure of the blob (related to $D_6$) will be
discussed in a future
paper. When the integration over the squared missing mass $M^2$ is performed,
the expression for the cross section
$\frac{d\sigma}{dt}= \int d M^2 \frac{d^2 \sigma}{dt dM^2}$
simplifies.
The result is illustrated in Fig.1b: the blob now stands for $D_6$ which
can directly be taken from ~\cite{BE}. It depends upon the angular
momentum variable $\omega$ which is conjugate to the total rapidity $Y$.

\begin{figure*}[tbh]
  \begin{center}
    \epsfig{file=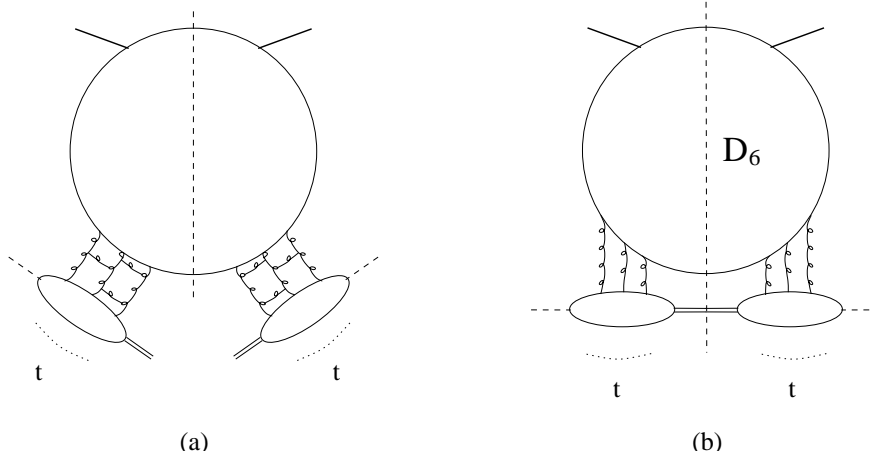,height=6cm} \bigskip
    \caption{Illustration of the process $\gamma^*+p \to \eta_c +X$.
     Diffractive cross section differential (a) and integrated (b) in the
     diffractive mass. }
\end{center}
\end{figure*}

The cross section can be written as
\be
\frac{d\sigma}{dt}=\xi\sum_{i=1,2} \int \frac{d\omega}{2 \pi i}
e^{Y\omega}
\int\frac {d\mu_1d\mu_2}
{\prod_{i=1}^6 k_i^2}
\Phi^i(1,2,3)[\Phi^i(4,5,6)]^*D_6(1,2,3,4,5,6;\omega) \, .
\label{gencrosssection}
\ee
Here $\Phi^i$ denotes the impact factor for the transitions
$\gamma^*\to\eta_c$ with the photon polarizations
$i=1,2$ and the color structure $d_{abc}$. $Y$ is the overall rapidity;
$t=-q^2$ is the invariant associated to the momentum transfer across
the impact
factor, and the arguments 1,2,3 and 4,5,6 refer to both the colour indices
$a_i$ and transverse momenta, $k_i$, of the gluons exchanged in the initial
amplitude $i=1,2,3$ and final (conjugated) amplitude $i=4,5,6$. Finally,
$d\mu_1$ is the integration measure for the 3 t-channel gluons on the lhs:
\be
d\mu_1=d^2k_1d^2k_2d^2k_3\delta^2(k_1+k_2+k_3-q) 
\ee
and $d\mu_2$ is the analogous integration measure for the three t-channel
gluons 4,5,6 on the rhs. The normalization factor $\xi$ will be
discussed in the next section.

From the analysis~\cite{BE} of the coupled integral equations it follows that
$D_6$ can be presented as a sum of different terms.
One term (denoted by $D_6^R$) is obtained by collecting the reggeizing
pieces: the outgoing six gluon state may contain configurations  where a
pair of two gluons is in an antisymmetric color octet configuration, which
satisfies the BFKL bootstrap condition and collapses into a single  gluon.
As a result, one obtains contributions with a smaller number of reggeized
gluons. It is convenient to separate these configurations from the rest,
i.e. to define the sub-amplitude $D_6^I$ which is 'irreducible' with respect
to this reduction procedure. This reduction leads
to the decomposition $D_6=D_6^R+ D_6^I$, separating the reggeizing (R) and
irreducible (I) parts. The $D_6^I$ term (eq.(6.3) of ~\cite{BE}) is rather
lengthy; however, for an odderon in  the (123) and (456) channels, we will
need, in the large-$N_c$ limit,  only one term, denoted by $W$, which
describes the transition of two reggeized gluons into six reggeized gluons:
all other terms will be shown (see section $4$) to be suppressed by a factor
of $1/N_c^2$ (or even higher powers of this). It is this piece of $D_6^I$
which yields the POO vertex.

Diagrammatically, the piece of $D_6^I$ which, in the large-$N_c$ limit
contains the Pomeron $\to$ odderon vertex has the structure shown in Fig.2.
\begin{figure*}[tbh]
  \begin{center}
    \epsfig{file=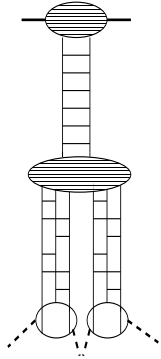,height=8cm,width=8cm,angle=-90} \bigskip
    \caption{Illustration of the process $\gamma^*+p \to \eta_c +X$:
     the blob in the center of the figure denotes the transition vertex
     Pomeron $\to$ 2 odderons.}
  \end{center}
\end{figure*}
The internal blob - with two gluons entering from above and six gluons
leaving below - defines the Pomeron $\to$ odderon (POO) vertex, and its
color structure is quite simple:
\begin{equation}
\delta _{b,b'} d_{a_1a_2a_3} d_{a_4 a_5a_6} W(1,2,3|4,5,6) \, ,
\label{vertex2to6}
\end{equation}
where the $b$, $b'$ are the color labels of the
reggeized gluons of the ladder above the POO vertex, $a_i$
the color indices of the reggeized gluons below the vertex
(counting from left to right).
The arguments of the function $W$ refer to the momenta of the gluons. Below
the POO vertex, we have the two noninteracting odderons:
the pairwise  interactions inside (123) and (456) lead to the color singlet
odderon Green's functions.
We note that this simple form emerges only after taking the large-$N_c$ limit.
In the more general case of finite $N_c$, the expression (\ref{vertex2to6})
has to be summed over permutations of the indices (123456). Moreover, in
Fig.2  below the POO vertex, we would have to include {\it all} pairwise
interactions  between the reggeized gluons. It is only in the large-$N_c$
limit that   any rung which connects the two color singlet (123) and (456)
costs a  suppression factor of the order $1/N_c^2$ and, therefore, can be
neglected.

The $D_6^R$ term is nothing but a sum of BFKL ladders in which, at the lower
end, the reggeized gluons split into two three or four elementary gluons.
Inserting this sum into the blob in Fig.1b and taking the large-$N_c$ limit,
we arrive at structures illustrated in Fig.3: the BFKL ladder couples to
odderon states consisting of three noninteracting gluons. Below we will
discuss this in further detail: starting from
the color structure of $D_6^R$, given in ~\cite{BE}, eq.(6.2),
it can be shown that all these contributions are subleading in $1/N_c$; in our
further discussion we will keep one of them which is suppressed by a factor
$1/N_c$ (all others are further suppressed).

\begin{figure}[htbp]
  \begin{center}
    \epsfig{file=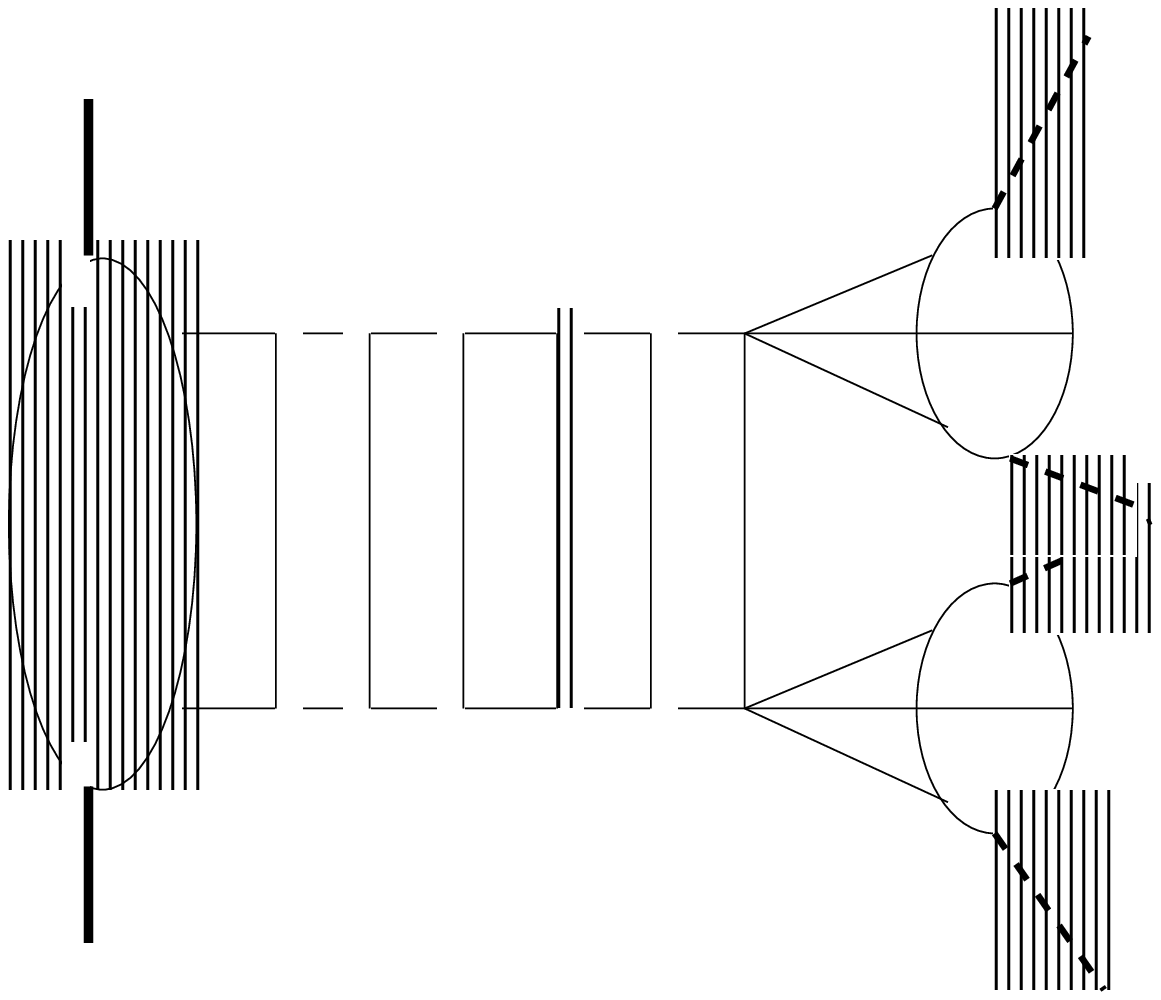,height=8cm,width=8cm,angle=-90} \bigskip
    \caption{The second contribution to the same process as Fig.1:
     the Pomeron couples to the two odderons which consist of three
     noninteracting gluons.}
  \end{center}
\end{figure}

All this discussion refers to Fig.1b which illustrates the {\it integrated}
inclusive cross section. In order to derive the pQCD formula for the
{\it differential} cross section, an alternative separation of $D_6$
is more suitable. For the case of $D_4$ which leads to the triple Pomeron
vertex, such a separation has been discussed in ~\cite{BW} and ~\cite{BV}:
as a result, a slightly different expression for the triple Pomeron emerges.
This question will be discussed in a forthcoming paper.

\section{Contribution of the reggeizing piece $D_6^R$}
Let us now analyze the two contribution to our inclusive cross section
in some detail. We start with the
ideologically (but not calculationally!) simpler part of the transition rate
corresponding to the reggeizing piece, $D_6^R$, of \cite{BE} (Fig. 3).
In the following, we will denote this piece by the superscript $P$.
The corresponding inclusive cross-section is given by the expression
\be
\frac{d\sigma^{(P)}}{dt}=\xi\sum_{i=1,2} \int \frac{d\omega}{2 \pi i}
e^{Y\omega}
\int\frac{d\mu_1d\mu_2}
{\prod_{i=1}^6 k_i^2} \Phi^i(1,2,3)[\Phi^i(4,5,6)]^*
D_6^R(1,2,3,4,5,6;\omega) \, .
\label{P1}
\ee
The function $D_6^R$ depending on all gluonic momenta is the reggeizing piece
of the  6-gluon amplitude found in \cite{BE}. It is given by the sum of BFKL
pomerons depending on various partial sums of the momenta of the
6 gluons 1,..6, multiplied by certain  colour factors.
All colour factors are obtained by permutations of inital (123)
or final (456) gluons from
\be
d^{a_1a_2a_3a_4a_5a_6}={\rm tr} (t^{a_1}t^{a_2}t^{a_3}t^{a_4}t^{a_5}t^{a_6})
+{\rm tr}(t^{a_6}t^{a_5}t^{a_4}t^{a_3}t^{a_2}t^{a_1}) \, ,
\ee
where $t^a$ is the quark colour matrix.
This evidently gives 8 different colour factors, so that $D^R$ contains
8 terms with different colour structure. Coupling $D^R$
to the two impact factors in (1) one has to contract its colour indeces
with a product $d_{a_1a_2a_3}d_{a_4a_5a_6}$ corresponding to the $C=-1$
exchange of three gluons. Then one finds that all colour factors in $D^R$
are transformed into the same common colour factor $F_c^{(R)}$
\be
d^{a_1a_2a_3a_4a_5a_6}d_{a_1a_2a_3}d_{a_4a_5a_6}=
\frac{(N_c^2-1)^2(N_c^2-4)^2}{8N_c^3}\equiv
N_c^5F_c^{(P)} \, .
\ee
At large $N_c$ it is $\sim N_c^5/8$, in correspondence with the general
rules of the $1/N_c$ expansion. Note however at $N_c=3$ its value
200/27$\sim 7.4$
is nearly 4 times smaller than given by the $N_c\to \infty$ limit 243/8
$\sim 30$.
Finally the overall factor $\xi$ is equal to
\be
\xi=\frac{1}{2} \frac{\pi}{16\pi^3}
\left(\frac{1}{4\cdot4(2\pi)^6(3!)}\right)^2 \, .
\label{xi}
\ee
The first factor $1/2$ corresponds to the averaging over the two photon
polarizations (we just consider transverse photon).
Then one has the standard $1/(16 \pi^3)$ phase space volume for the
diffractive process times a $\pi$ factor due to angular averaging.
In the squared term there is the contribution of $(2\pi)^{-4}$ absent in each
$d\mu$ and an extra $(2\pi)^{-2}$ which we associate to the
impact factor \cite{Czyzewski:1997bv} in our normalization.
Moreover one has the symmetry factor $1/3!$ for each of the gluon
triplets, 1/4 from the colour factor which is in fact $(1/4)d_{abc}$ and 1/4
from the integrations over $s_i$, instead of over $k_{i-}$ in the definition
of the impact factor.

Separating the common  factor $g_s^4N_c^5F_c^{(P)}$, where $g_s$ is the
strong coupling constant, we shall reproduce the rest of  the amplitude $D^R$
from \cite{BE} in a simplified manner, taking into account that, first, in
the pomeron of Fig. 3 the total momentum of the two gluons is zero and, second,
in the high-energy limit the amplitude for the leading contribution is
symmetric in the two gluons. Thus this ampitude $P(k)$
(amputated, that is, without external gluon propagators) depends only on one
of the gluon momenta. In terms of $P(k)$ the reggeizing piece is then
given by a sum of 31 terms:
\be
D_6^R=\sum_{i=1}^6P(i)-\sum_{i\ne k=1}^3P(ik)-\sum_{i\ne k=4}^6P(ik)
-\sum_{i=1}^3\sum_{k=4}^6P(ik)+\sum_{i\ne k=1}^3\sum_{l=4}^6P(ikl)+
P(q) \, .
\label{DR}
\ee
Here the notations $il$ and $ilm$ denote sums of gluon momenta $k_i+k_l$ and
$k_i+k_l+k_m$ respectively.
Expression (\ref{DR}) can further be simplified if we take into account
that in (\ref{P1})
the integration over all momenta is done for a function which is totally
symmetric in the gluons (123) and (456). Moreover, it is symmetric
under the interchange $(123)\leftrightarrow (456)$. Therefore, on the rhs
of (\ref{DR}) the terms inside each sum are identical, and we get
\be
D_6^R=6P(1)-6P(12)-9P(14)+9P(124)+P(q).
\ee

The $\gamma^*\to \eta_c$ impact factor is given by \cite{Czyzewski:1997bv}
\be
\Phi^i(1,2,3)=b\epsilon_{ij}\frac{q_j}{q^2}\phi(1,2,3),\ \  i,j=1,2.
\label{impfact}
\ee
Here
\be
\phi(1,2,3)=\sum_{i=1}^3\frac{q(q-2k_i)}{M^2+(q-2k_i)^2}-
\frac{q^2}{M^2+q^2}
\ee
and $M^2=Q^2+4m_c^2$, where
$Q^2$  is the photon virtuality and $m_c$ the charmed quark mass.
The coefficient $b$ is given by
\be
b=\frac{16}{\pi}e_cg_s^3\frac{1}{2}m_{\eta_c} b_0 \, ,
\ee
where $e_c=(2/3)e$ is the electric charge of the charmed quark
$m_{\eta_c}$ is the $\eta_c$ meson mass  and
$b_0$ can be determined from the known radiative width
 $\Gamma(\eta_c \to \gamma\gamma)$= 7 KeV:
\be
b_0=\frac{16\pi^3}{3e_c^2}\sqrt{\frac{\pi\Gamma}{m_{\eta_c}}} \, .
\ee
The impact factor (\ref{impfact}) is symmetric in the three gluon momenta, and
it vanishes if any of the momenta goes to zero. Taking in (4) the product of
the two impact factors and summing over
polarizations we obtain
\be
F_{\gamma^*\to \eta_c} \phi(1,2,3) \phi(4,5,6) \, ,
\ee
where
\be
F_{\gamma^*\to \eta_c}=\frac{b^2}{q^2} \, .
\ee

The pomeron $P(k)$ can be presented as a convolution of the BFKL Green
function with the colour distribution $\rho(r)$ in the hadronic target
\be
P(k)=-g_s^2\int d^2r'G(Y,k,r')\rho(r').
\label{evolprot}
\ee
Note the minus sign. After transforming the initial momentum space expression
for the impact factor into the coordinate space, the impact factor
is proportional to $1-\exp(ikr)$. When multiplying with the Pomeron
Green's function and doing the $k$-integral, there is no contribution from
the `1' (since, in coordinate space, the Pomeron Green's function vanishes
when both arguments coincide), and the nonzero contribution comes from the
second term, $-\exp(ikr)$.
The Green function has to be taken in a mixed representation, momentum $k$ at
the odderon side, coordinate $r'$ at the proton side. Also,
one side of the Green's function is amputated, the other not.
\be
G(Y,k,r')=-\frac{1}{8\pi^2}qr'\int \frac{d\nu}{\nu^2+1/4} e^{Y\omega(\nu,0)}
\left(\frac{qr'}{2}\right)^{2i\nu}
\label{greenfP}
\ee
with
\be
\omega(\nu,n)=2\bar{\alpha}_s
\Big(\psi(1)-{\rm Re}\,\psi(\frac{1+|n|}{2}+i\nu)
\Big),\ \ \bar{\alpha}_s=\frac{\alpha_sN_c}{\pi}.
\label{intercept}
\ee
At large rapidities small $\nu$'s dominate.
Due to the finite dimension of the target, in (\ref{greenfP} the values of
$r'$ are limited by the radius $R$. So we can neglect factor the
$(r')^{2i\nu}$, and the integration over
$r'$ will be replaced by the average transverse dimension of the target $R$.
So at high $Y$  the Green function  gives  a factor
\be
-\frac{1}{2\pi^2}kR\sqrt\frac{\pi}{aY}e^{\Delta Y}
\exp\left(-\frac{\ln^2 kR}{4aY}\right).
\label{fact1P}
\ee
Here $\Delta=\omega(0,0)=\bar{\alpha}_s\,4\ln 2$ is the BFKL intercept
with $a=7\bar{\alpha}_s\zeta(3)$.
The second exponential factor cuts the integration over $k$ to values
$\ln^2k<aY$. However, since the integration in (\ref{P1}) is in fact
convergent, we may drop this factor at high enough $Y$.
Putting (\ref{fact1P}) into (\ref{DR}) we get for the cross-section
\be
\frac{d\sigma^{(P)}}{dt}=\xi g_s^6\frac{b^2}{2\pi^2q^2M^3}N_c^5
F_c^{(P)} R e^{\Delta Y}\sqrt\frac{\pi}{aY}I\left(\frac{q}{M}\right),
\label{cross1P}
\ee
where the dimensionless function $I(q/M)$ is given by the integral
\be
I\left(\frac{q}{M}\right)=M^3\int\frac{d\mu_1
d\mu_2}{\prod_{i=1}^6k_i^2}\phi(1,2,3)\phi(4,5,6)
\Big(6|k_1|-6|q-k_1|-9|k_1-k_4|+9|q-k_1-k_4|+q\Big).
\label{I8}
\ee
The cross section (\ref{cross1P}) is of the order $\alpha_s(\alpha_s N_c)^5$.
The 8-dimensional integral  (\ref{I8}) is non-factorizable and can be done
only numerically.

As an alternative way of evaluating this contribution to the cross section,
one might try to factorize the integration over gluonic momenta and to
transform the pomeron amplitude to the coordinate space using
\be
k^{1+2i\nu}=-(1+4\nu^2)
\int\frac{ d^2r}{2\pi r^3}\left(\frac{2}{r}\right)^{2i\nu}e^{ikr} \, .
\ee
Taking the limit $\nu\to 0$ and introducing a function of $r$
\be
h(r)=\int\frac{d\mu_1\phi(1,2,3)}{k_1^2k_2^2k_3^2}e^{ik_1r} \, ,
\ee
we find the integral $I$ as
\[
I\left(\frac{q}{M}\right)=-M^3\int \frac{d^2r}{2\pi r^3}\]\be
\Big\{\Big(6h(r)h(0)-6e^{iqr}h(-r)h(0)-
9h(r)h(-r)+9e^{iqr}h^2(-r)+e^{iqr}h^2(0)\Big)-\Big( r=0\Big)\Big\}.
\ee
In obtaining this expression we used the fact that the change $q\to -q$ is
equivalent to changing $r\to -r$ in $h(r)$. Also, we have taken into
account that $P(k=0)=0$, in order to subtract the value of the  brackets at
$r=0$ and thus to improve the convergence of the integral at $r=0$.
Passing to the function
\be h_1(r)=h(r)-e^{iqr}h(-r),\ \ h_1(0)=0,
\ee
we can rewrite (23) in a simpler form, which also shows that the right-hand
side is real:
\be
I\left(\frac{q}{M}\right)=-M^3
\int \frac{d^2r}{2\pi r^3}\Big(6h(0){\rm Re}\,h_1(r)-\frac{9}{2}|h_1(r)|^2
-(1-\cos (qr))h^2(0)\Big).
\ee
Now we have only 6 integrations to be done numerically.
However the presence of the oscillating factor
makes such calculations very difficult. We therefore use Eq. (\ref{I8}) and performed the integrations by Monte-Carlo methods.
The results will be presented in Sec.6, together with the contribution from
the POO vertex.

\section{The POO vertex in leading order in $N_c$}
Next let us consider the irreducible part $D_I$ of the amplitude for
6 reggeized gluons. Its contribution to the cross section will be denoted
by the superscript POO. Starting from \cite{BE} (eq.(6.3), we
note that the rhs satisfies a BFKL-like equation which we
write in the symbolic form
\be
(H_6-E)D_I=D_I^{(0)} \, .
\label{evol6}
\ee
Here $H_6$ is the total Hamiltonian for 6 reggeized gluons, which is
a sum of pairwise interactions and of gluon trajectories and describes
their evolution, without changing the number of gluons. The energy
$E=1-j=-\omega$ is just one minus the intercept. The driving term
of the equation is a sum of  terms which describe transitions with a change
of the number of gluons, from "irreducible" configurations of
2, 4, or 5 gluons to 6 gluons.
At this moment it is important to invoke the approximation of large
number of colours $N_c\to\infty$. In this approximation, any interaction
inside the outgoing six gluon state which connects
colourless groups of gluons is damped by
$1/N_c^2$ and can be neglected. This means, in particular, that once a
pair of states with colour color structure of two odderons is formed
in the driving term, $H_6$ in Eq.~(\ref{evol6}) contains no further
interaction between these two odderon states. All what $H_6$ does is
to build up the bound states of the gluons (123) and (456).
So in order to find the terms relevant for the POO transitions
we only have to see whether the final two 0dderon states couples to the
driving term. One immediately sees that, at large $N_c$, the irreducible
configurations of 4 and 5 gluons, $D_4^I$ and $D_5^I$, reduce to the
splitting of the initial
pomeron into two pomerons and thus cannot couple to the
two odderon final state. Therefore, the transitions of interest
can only occur in terms which describe transitions of
2 gluons to 6 gluons. In \cite{BE} 4 such terms of different
colour structure were found.

The first group is given by a sum
\be
 d_{a_1a_2a_3}d_{a_4a_5a_6}W(1,2,3|4,5,6)+
 d_{a_1a_2a_4}d_{a_3a_5a_6}W(1,2,4|3,5,6)+ ...,
\label{first}
\ee
where the sum extends over all (ten) partitions of the six gluons into two
groups containing three gluons each. Projecting onto the two
odderon colour state we find the colour factor for the first term in
(\ref{first}) \be
d_{a_1a_2a_3}d_{a_4a_5a_6}d_{a_1a_2a_3}d_{a_4a_5a_6}=
\frac{(N_c^2-1)^2(N_c^2-4)^2}{N_c^2}\equiv N_c^6F_c^{(POO)}\sim N_c^6,
\label{first1}
\ee
whereas for all the rest terms we have
\be
d_{a_1a_2a_4}d_{a_3a_5a_6}d_{a_1a_2a_3}d_{a_4a_5a_6}=
\frac{(N_c^2-1)(N_c^2-4)^2}{N_c^2}\sim N_c^4.
\ee
So at large $N_c$ we will retain only the first term in the sum, (\ref{first}).

Apart from the $W$ terms, the remaining driving terms in (6.3) of
(\ref{evol6}) with transitions
from 2 to 6 gluons contain 3 more groups of terms of different colour
structure. Terms denoted by $L$ in \cite{BE} are given by a sum
\be
 f_{a_1a_2a_3}f_{a_4a_5a_6}L(1,2,3|4,5,6)+
 f_{a_1a_2a_4}f_{a_3a_5a_6}L(1,2,4|3,5,6)+ ...,
\ee
with the sum, again, extending
over all (ten) partitions of the six gluons into two
groups containing three gluons each, and the function being described
in ~\cite{BE}. Obviously
these terms give zero when projeted onto the colour state of two odderons.

Finally, the terms denoted by $I$ and $J$ are given by  sums
\be
d^{a_1a_2a_3a_4}\delta_{a_5a_6}I(1,2,3,4|5,6)+
d^{a_1a_2a_3a_5}\delta_{a_4a_6}I(1,2,3,5|4,6)+...
\ee
and
\be
d^{a_2a_1a_3a_4}\delta_{a_5a_6}I(1,2,3,4|5,6)+
d^{a_2a_1a_3a_5}\delta_{a_4a_6}I(1,2,3,5|4,6)+...,
\ee
with the sum extending over all partitions of 6 gluons into two groups
with 4 and 2 gluons,known function $I$ and $J$ and
\be
d^{a_1a_2a_3a_4}={\rm tr}(t^{a_1}t^{a_2}t^{a_3}t^{a_4})+
{\rm tr} (t^{a_4}t^{a_3}t^{a_2}t^{a_1}).
\ee
Projecting onto the colour state of two odderons we find
non-zero colour factors of the type
\be
d^{a_1a_2a_4a_5}\delta{a_3a_6}d_{a_1a_2a_3}d_{a_4a_5a_6}=
\frac{(N_c^2-1)(N_c^2-4)^2}{4N_c^2}\sim N_c^4 \, .
\ee
So although these terms seem to involve transitions POO, they are
down by a factor $N_c^2$ as compared to (\ref{first1}).

So in the end we find that, in the large $N_c$ limit, the transitions POO are
fully described by the function $W(1,2,3|4,5,6)$
which represents a convolution of the pomeron with a POO vertex.
The functional form of $W$ in ~\cite{BE} is rather complicated, however
a closer inspection shows a surprisingly simple structure
\cite{Bartels:2000yt}. Let us briefly recapitulate this structure.
It is convenient to introduce
two operators which transform a function of momenta of two
gluons into a new function which depends upon momenta of three gluons.
Namely, define
$\hat{S}$ to be an operator acting on 2-gluon
states and with values on the 3-gluon states, which performs an
antisimmetrization in the 2 incoming gluons, splits the first of them
in two and sums over the cyclic permutations of the outgoing gluons:
\be
\hat{S}(1,2,3|1',2')\phi(1',2')=\frac{1}{2}\sum_{(123)}
[\phi(12,3)-\phi(23,1)] \, .
\ee
Next define another operator $\hat{P}$ which
performs an antisimmetrization in the 2 incoming gluons and splits the first
of them  in three outgoing gluons:
\beq
\hat{P}(1,2,3|1',2'))\phi(1',2')
= \frac{1}{2}  [\phi(123,0))-\phi(0,123)] \, .
\eeq
Apart from these operators we introduce a function $f(1,2|3,4)$,
antisymmetric in the first and second pairs of gluons and symmetric
under the interchange $(12)\leftrightarrow (34)$, as a sum of
functions $G(1,2,3)$, which were introduced in \cite{BW,BV} in the context
of the three-pomeron vertex:
\be
f(1,2|3,4)=G(1,23,4)-G(2,13,4)-G(1,24,3)+G(2,14,3) \, .
\label{fdef}
\ee
The explicit form of the general function $G(1,2,3)$ is
not important for our purpose (it can be found e.g. in \cite{BW,BV}).
We only have to know that
\be
G(1,2,3)=G(3,2,1), \ \ G(0,2,3)=G(1,2,0)=0\, ,
\label{prop1}
\ee
and that, up to a coefficient, $G(1,0,3)$ is given by the BFKL Hamiltonian
$H_2$ applied to the pomeron:
\be
G(1,0,3)=-\frac{1}{N_c}\Big(H_2 P\Big)(1,3)\, .
\label{prop2}
\ee

In terms of $\hat{S}$, $\hat{P}$ and $f$ we find
\be
W(1,2,3|4,5,6)=-\frac{1}{8}g_s^4(\hat{S}_1-\hat{P}_1)
f_{12}(\hat{S}_2^{\dagger}-\hat{P}_2^{\dagger})\, ,
\label{vertex}
\ee
where the indices 1 and 2 refer to the triplets of gluons (123) and (456),
respectively.

\section{Part of the cross-section with a POO transition}
To find the cross-section corresponding to the $\eta_c$ production via the
POO transition (Fig. 1) we have to couple the POO vertex with the two odderons
attached to the initial and final $\gamma^*\to \eta_c$ impact factors.
To write it in a compact form we introduce the Green functions
$G_3^{(1)}$ and $G_3^{(2)}$ for the initial and final odderons
composed of gluons 123 and 456 respectively. They evolve the odderon state
in the rapidity interval with length $y<Y$.
Then using (\ref{vertex}) we can write the cross-section as
\be
\frac{d\sigma^{(POO)}}{dt}=-\frac{1}{8}g_s^4N_c^6F_c^{(POO)}
\xi\sum_{i=1,2} \int dy <\Phi^i_1|G_3^{(1)}(\hat{S}_1-
\hat{P}_1)f_{12}(\hat{S}_2^{\dagger}-\hat{P}_2^{\dagger})
G_3^{(2)}|\Phi_2^i> \, ,
\ee
where it is assumed that the averaging is done independently
over the gluons 123 (index 1) and 456 (index 2), function $f$ depending
on both groups of variables.

At this point we recall that the full set of odderon states
and consequently the full Green function $G_3$ are unknown. We are going to
use a part of it corresponding to the solutions found in \cite{Bartels:2000yt},
which have the maximal intercept of all known states and besides have a
non-zero coupling to the perturbative $\gamma^* \to \eta_c$ impact factor.
These odderon states are expressed via the known antisymmetric pomeron states
$E^{(\nu,n)}$ with odd values of $n$:
\beq
\Psi^{(\nu,n)}(1,2,3)=c(\nu,n)
\frac{1}{k_1^2k_2^2k_3^2 }
\hat{S}(1,2,3|1',2'){k'_1}^2{k'_2}^2E^{(\nu,n)}(1',2') \, ,
\label{oddstate}
\ee
where
\be
c(\nu,n)=\sqrt{\frac{g_s^2 N_c}{-3(2\pi)^3\omega(\nu,n)}}
\ee
and $\omega(\nu,n)$ is given by Eq. (\ref{intercept}).
The part of the Green function $G_3$ corresponding to these states
then aquires a form similar to the pomeron Green function \cite{BBCV}
\beq
G_3(y_1|1,2,3|1',2',3')=
\sum_{{\rm odd}\ n}\int d\nu e^{y_1\, \omega(\nu,n)}\beta(\nu,n)
\Psi^{(\nu,n)}(1,2,3)
{\Psi^{(\nu,n)}}^*(1',2',3') \, ,
\label{greenf}
\eeq
with
\be
\beta(\nu,n)=\frac{(2\pi)^2(\nu^2+n^2/4)}{[\nu^2+(n-1)^2/4][\nu^2+(n+1)^2/4]}
\label{weight}
\ee
and $\Psi^{(\nu,n)}$ given by (\ref{oddstate}).

Now we use the fact that we only study our cross-section in the region where
both the rapidity of the pomeron $y$ and that of the odderon $y_1=Y-y$
are large. This allows to retain in (\ref{greenf}) only the branch $|n|=1$
with a maximal
intercept and also restrict the integration over $\nu$ to small values.
In the limit $\nu\to 0$ the coupling of the odderon to $\gamma^*\to \eta_c$
impact factor was calculated in \cite{BBCV} to give
\be
\langle \Phi_{\gamma \to \eta_c}^i |\Psi^{(\nu,n)} \rangle
= - \frac{i}{\pi}b\epsilon_{ij} \frac{q_j}{q} \frac{1}{c(\nu,n)}
 q^{2i \nu}  \frac{1}{q^2+M^2} \, .
\label{scalarodd}
\ee
 After summation over the photon polarizations the
two matrix elements (\ref{scalarodd})  provide a factor
\be
\frac{1}{\pi^2}\frac{b^2}{c(\nu_1,n_1)c(\nu_2,n_2)}
\frac{1}{(q^2+M^2)^2}q^{2i(\nu_1-\nu_2)} \, ,
\label{summedpol}
\ee
where $(\nu_1,n_1)$ and $(\nu_2,n_2)$ refer to the summation and
integration variables in $G_3^{(1)}$ and $G_3^{(2)}$ respectively.
At finite value of $q$ one can neglect the last factor in (\ref{summedpol}).

We are left with the matrix element
\be
T\equiv <\Psi_1^{(\nu_1n_1)}|(\hat{S}_1-
\hat{P}_1)f_{12}(\hat{S}_2^{\dagger}-\hat{P}_2^{\dagger})
|\Phi_2^{(\nu_2,n_2)}> \, .
\label{matel}
\ee
Its calculation is greatly simplified by the property of the
odderon state  (\ref{oddstate}) found in \cite{Bartels:2000yt}.
For any function $\phi(1,2)$ of two gluon momenta and odderon state
(\ref{oddstate}) one has
\be
<\Psi^{(\nu,n)}|(\hat{S}-\hat{P})|\phi>=
 \frac{1}{c(\nu,n)}<E^{(\nu,n)}|\phi> \, .
\label{simpli}
\ee
Note that the matrix element on the left-hand side is taken in the space
of three gluons, whereas that on the right-hand side is taken in the space
of only two gluon momenta. This property greatly simplifies the matrix
element (\ref{matel}). Using (\ref{simpli}) we get for it
\be
T=\frac{1}{c(\nu_1,n_1)c(\nu_2,n_2)}
<E^{(\nu_1,n_1)}_1|f_{12}|E^{(\nu_2,n_2)}_2> \, .
\label{Tsimpli}
\ee

Now we again use the fact that $y_1$ is large and so only values
$|n|=1$ and $|\nu |<<1$ contribute.
At $|n|=1$ and small $\nu$ the pomeron wave funtions entering (\ref{Tsimpli})
reduce to $\delta$ functions of gluon momenta \cite{BBCV}
\be
E^{(\nu,\pm 1)}(1,2)_{\nu\to 0}=\frac{i}{2\pi q}
\Big(\delta^2(k_1)-\delta^2(k_2)\Big),\ \
E^{(\nu,\pm 1)}(4,3)_{\nu\to 0}=\frac{i}{2\pi q}
\Big(\delta^2(k_4)-\delta^2(k_3)\Big) \, .
\ee
Note that the second wave function has to be taken conjugate.
Putting this into (\ref{Tsimpli}) and recalling (\ref{fdef}) and the
properties of the function $G$ (Eq. (\ref{prop1}) and (\ref{prop2}) we get
\be
T=-\frac{1}{c(\nu_1,n_1)c(\nu_2,n_2)}\frac{4}{(2\pi)^2q^2}
G(q,0,-q)=
\frac{1}{c(\nu_1,n_1)c(\nu_2,n_2)}\frac{1}{\pi^2q^2}\frac{1}{N_c}
(H_2P)(q) \, .
\ee
The last factor is just the BFKL Hamiltonian applied to the Pomeron state.

As in Sec. 2, to find the pomeron $P(q)$ attached to the hadronic target
 we present it as the BFKL Green function applied to
the colour distribution in the hadron, Eq. (\ref{evolprot}). Again we need a
mixed amputated-non-amputated Green function in the momentum space in
its amputated part, Eq. (\ref{greenfP}).
Applying the BFKL Hamiltonian we get
\be
H_2G(y,q,r)=\frac{1}{8\pi^2}qr\int d\nu
\omega(\nu,0)
e^{y\omega(\nu,0)} \frac{2^{-2i\nu}(qr)^{2i\nu}}{\nu^2+1/4} \, .
\ee
At large $y$ with finite $q$ and $r$ we neglect all dependence on $\nu$
except in the exponential to obtain similarly to (\ref{fact1P})
\be
HG(y,q,r)=\frac{1}{2\pi^2}qr\Delta e^{y\Delta}
\sqrt{\frac{\pi}{ay}} \, .
\ee
Integration of $r$ with the target colour density converts it into
the average target transverse radius $R$ with a minus sign ,
see (\ref{evolprot})).

So we find for the matrix element $T$
\be
T=-\frac{1}{2\pi^4q^2}\frac{1}{c(\nu_1,n_1)c(\nu_2,n_2)}
\frac{g_s^2}{N_c}qR\Delta e^{y\Delta}
\sqrt{\frac{\pi}{ay}} \, .
\label{matel2}
\ee
We have finally to do the integrations over $\nu_1$ and $\nu_2$
and summations over $n_1$ and $n_2$, which in the limit of high $y_1$ only
take values $\pm 1$. These latter  summations  are trivial and give a
factor 4.  At $|n|=1$ and small $\nu$ we have
\be
\omega(\nu,\pm 1)=-2\bar{\alpha}_s\zeta(3)\nu^2 \, .
\ee
One of the denominators in (\ref{weight}) reduces to $\nu^2$ and is singular at
$\nu\to 0$. However this singularity  is cancelled by the square of
the  $1/c^2(\nu,\pm 1)$ coming from (\ref{summedpol}) and
(\ref{matel2}).
Therefore we find at small $\nu$
\be
\frac{\beta(\nu,\pm 1)}{c^2(\nu,\pm 1)}=12\pi^3\zeta(3) \, .
\ee
Neglecting all the rest $\nu$-dependence except in the
exponential in (\ref{greenf}), integrations over $\nu_1$ and $\nu_2$ 
provide a factor
\be
\frac{\pi}{2\bar{\alpha}_s\zeta(3)y_1} \, .
\ee

Combining all the factors we finally get for the cross-section a
simple expression
\be
\frac{d\sigma^{(POO)}}{dt}= 18 \pi \xi N_c^6 F_c^{(POO)} \int dy \,
g_s^6\frac{\Delta}{N_c}\frac{b^2R\zeta(3)}{q(q^2+M^2)^2}
e^{\Delta y}\frac{1}{\bar{\alpha}_sy_1}\sqrt{\frac{\pi}{ay}} \, .
\ee
It steadily grows as $q^2$ diminishes and behaves as $1/q$ at small $q$.
Integrating over all $q$ we find the cross-section
\be
\sigma^{(POO)}=9 \pi^3 \xi N_c^6 F_c^{(POO)} \int dy \,
g_s^6 \frac{\Delta}{N_c}
\frac{b^2R\zeta(3)}{M^3}e^{\Delta y}\frac{1}{\bar{\alpha}_sy_1}
\sqrt{\frac{\pi}{ay}} \, .
\ee
It  falls with $Q^2$ and the meson mass as $1/(Q^2+m_{PS})^{3/2}$.
The cross-section has an order  $\alpha_s(\alpha_sN_c)^6$,
an order higher in $\alpha_s N_c$ than the leading contribution
given by the reggeizing part (Fig. 2). This implies that
the vertex POO has the same order $\alpha_sN_c$ as the triple pomeron
vertex.

\section{Numerical results}
Both the cross-section with a pure pomeron exchange and with a POO vertex
have a simple dependence on energies, which separates into a  factor
with the expected behaviour at large rapidities. Separating also all
the rest non-trivial factors we find the pure pomeron exchange
contribution as
\be
\frac{d\sigma^{(P)}}{dt}= c^{(P)}\alpha_{em}\alpha_s(\bar{\alpha}_s)^5
b_0^2\frac{m_{\eta_c}^2R}{q^2M^3}
I\left(\frac{q}{M}\right)f^{(P)}(Y) \, ,
\label{diffP}
\ee
where
\beq
f^{(P)}(Y)=e^{\Delta Y}\sqrt\frac{\pi}{aY}
\eeq
and
\be
c^{(P)}=\frac{F_c^{(P)}}{5184\,\pi^6} \, .
\ee
The part originating from the POO vertex is
\be
\frac{d\sigma^{(POO)}}{dt}= c^{(POO)}
\alpha_{em}\alpha_s \int dy \, (\bar{\alpha}_s)^6
b_0^2\frac{m_{\eta_c}^2R}{q(q^2+M^2)^2}f^{(POO)}(Y,y) \, ,
\label{diffPOO}
\ee
where now
\beq
f^{(POO)}(Y,y)=\frac{1}{\bar{\alpha}_s(Y-y)}e^{\Delta y}\sqrt\frac{\pi}{ay}
\eeq
and
\be
c^{(POO)}=\frac{F_c^{(POO)}\zeta(3)\ln 2}{36 \pi^4} \, .
\ee
The cross-section with the POO vertex integrated over all transferred
momenta is
\be
\sigma^{(POO)}=\frac{1}{2}\pi
c^{(POO)}\alpha_{em}\alpha_s \int dy \, (\bar{\alpha}_s)^6
b_0^2\frac{m_{\eta_c}^2R}{M^3}f^{(POO)}(Y,y) \, .
\ee
With $N_c=3$ the colour factors become
\be
F_c^{(P)}=200\cdot 3^{-8},\ \ F_c^{(POO)}=1600\cdot 3^{-8} \, .
\ee
Regarding the region of integration in rapidity $y$, as explained in the
introduction, we consider the interval $\delta Y < y < Y-\delta Y$
to warrant the use of the asymptotic forms for both the pomeron and the
odderons. We choose $\delta Y=3$.

One should also take a certain care with the coupling constants in the
expressions for the cross-sections. In fact they refer to different scales
relevant to the studied processes. Obviously one of the coupling
constants refers to the coupling to the
proton at a  small and so non-perturbative scale. For the process
mediated by the pure pomeron all other coupling constants
are to be taken at the scale of the $\gamma^*\to\eta_c$ transition,
that is the maximal of $q$ and $M$. The cross-section falls quite
rapidly as $q$ becomes larger than $M$, so that we can safely take $M$
as the relevant scale. For the process with the POO transition however
only three of the remaining 6 $\alpha$'s refer to this scale. Other three
are to be taken at an intermediate scale, characteristic for the
POO transition at rapidity $y$. For  high rapidities $Y$ and $y$
one can use the fact that  the characteristic momenta $k$
in the BFKL pomeron at rapidity $Y$ have the order $\ln k\sim\sqrt{Y}$.
Then one obtains a crude estimate for the coupling constant at the
POO junction as
\be
\alpha_{POO}\sim\sqrt{\frac{Y}{y}}\alpha_s(M) \, .
\label{alphasPOO}
\ee
We have taken $Y=\ln(1/x)$ with $x$ defined as
\beq
x=\frac{m_{\eta_c}^2+Q^2}{s+Q^2} \, .
\eeq

Passing to concrete values of the coupling constants
we take $\alpha_s(M)$ as given by the leading order $\beta$-function with
3 or 4 flavours $N_f$ and $\Lambda_{QCD}=0.2$ GeV/c. The value of $\alpha_s$ at the
POO junction was taken according to (\ref{alphasPOO}). As to the values
of the
pomeron intercept and its coupling to the proton,
we have borrowed them from ~\cite{AB},
where  the proton structure function at small $x$ was fitted by the pomeron
exchange. From this fit one extracts both $\Delta$ and the product
$\alpha_sR$:
\beq
\Delta=0.377,\ \ \alpha_sR=0.096\ {\rm fm},\ \ (N_f=3),\ \
\alpha_sR=0.058\ {\rm fm}\ \ (N_f=4) \, .
\eeq
At first sight one may expect a large difference in the results
for different  $N_f$.
However a smaller
value of $\alpha_sR$ for $N_f=4$ is compensated by a larger value of
the rest of the coupling constants, so that the final results are
practically independent of the number of flavours
taken into account (see Fig. 7 below).

The calculation of the POO contribution (\ref{diffPOO}) is straightforward.
To find the contribution from the pure pomeron exchange (\ref{diffP})
one has to calculate integral (\ref{I8}). We did this using the standard
Monte-Carlo program VEGAS. The results for $I(q/M)$ are presented in Fig. 4.

\begin{figure}[htbp]
  \begin{center}
    \epsfig{file=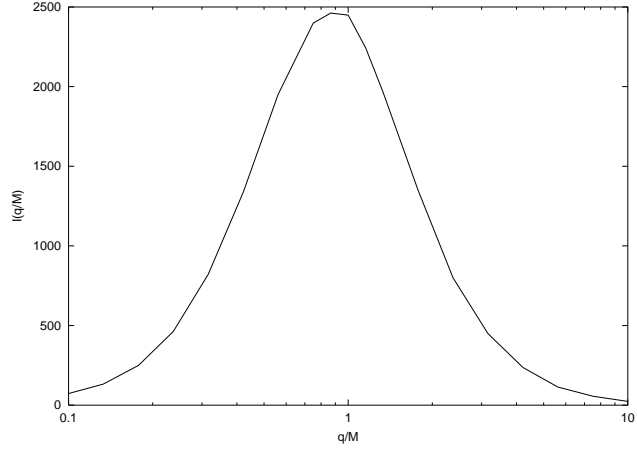,height=6cm} \bigskip
    \caption{The function $I(q/M)$ from eq.(\ref{I8}).}
  \end{center}
\end{figure}

Our final results for the differential cross-sections (\ref{diffP})
and (\ref{diffPOO})
and their sum (for $N_f=3$) are presented in Figs. 5 and 6  for
$\sqrt{s}=300$ GeV and $Q^2=0$ and 25 (GeV/c)$^2$ respectively.

\begin{figure}[htbp]
  \begin{center}
    \epsfig{file=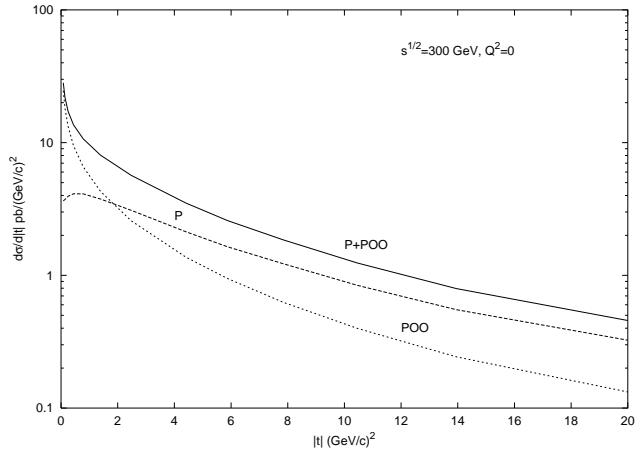,height=6cm} \bigskip
    \caption{Differential cross-sections $d\sigma/dt$ from the
    reggeized pomeron exchange (P),
    POO transition (POO) and total at $Q^2=0$.}
  \end{center}
\end{figure}

\begin{figure}[htbp]
  \begin{center}
    \epsfig{file=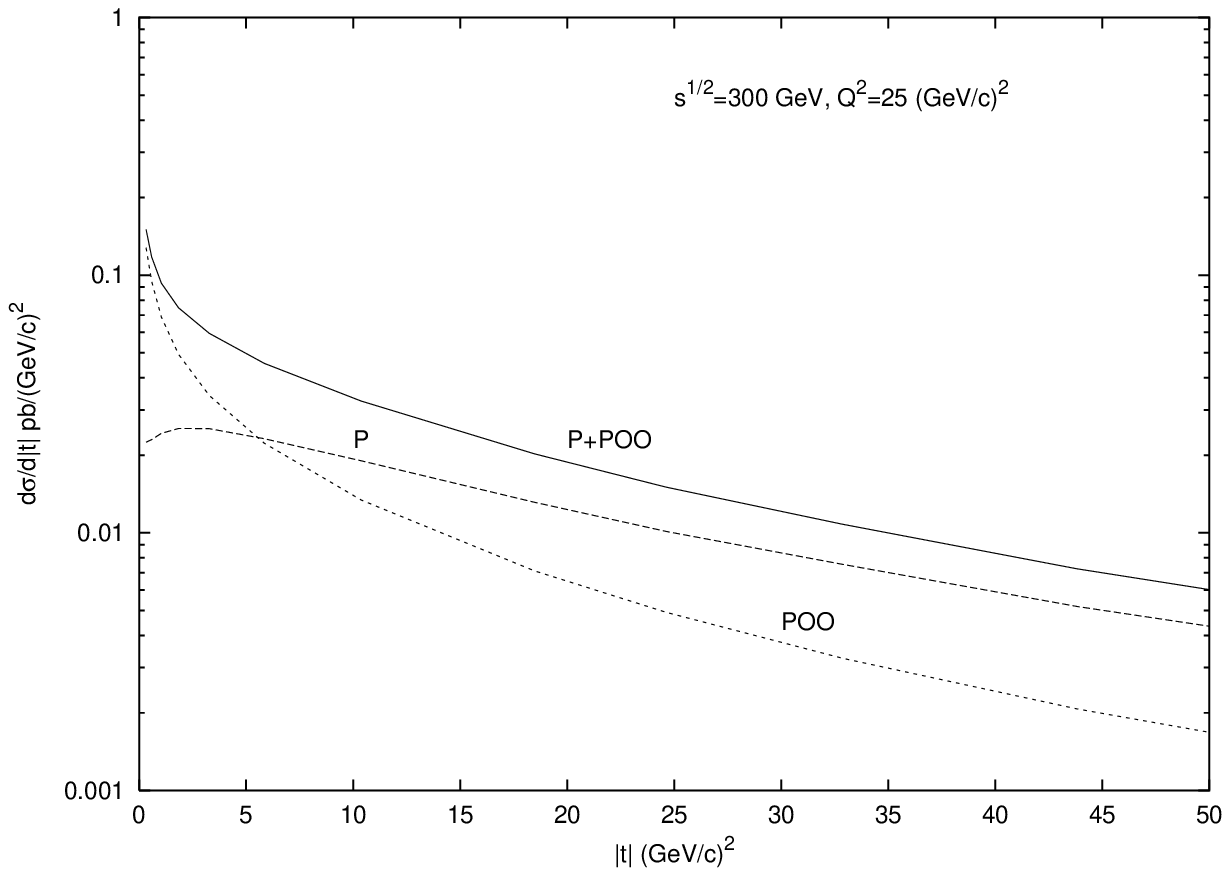,height=6cm} \bigskip
    \caption{Differential cross-sections $d\sigma/dt$ from the
    reggeized pomeron exchange (P), POO transition (POO) and total at
    $Q^2=$25 (GeV/c)$^2$}
\end{center}
\end{figure}

As we see, the contribution from the POO transition turns out  to
be of the same order as the one coming from the direct coupling of the
pomeron to non-interacting gluons (P contribution).
However the two contributions
seem to behave differently at small transferred momenta.
From (\ref{diffPOO}) we see that at $q\to 0$ the POO contribution rises as
$1/q$. On the other hand, the  P contribution does not
show such a behaviour and seems to tend  to a constant or zero at small $q$.
This may help to see  the POO contribution against the reggeizing
pomeron contribution at very low $q$.

Integrated over  transferred momenta the total cross-sections are found
at $Q^2=0$ to be
\[
\sigma^{(P)}=34\ {\rm pb},\ \ \ \sigma^{(POO)}=31\ {\rm pb},\ \
\sigma=\sigma^{(P)}+\sigma^{(POO)}=65\ {\rm pb}
\] and at $Q^2=$25 (GeV/c)$^2$
\[
\sigma^{(P)}=0.87\ {\rm pb},\ \ \ \sigma^{(POO)}=0.67 \ {\rm pb},\ \
\sigma=\sigma^{(P)}+\sigma^{(POO)}=1.54\ {\rm pb} \, .
 \]

The total integrated cross sections $\sigma$ (sum of P and POO)
in the range of photon
virtualities $0< Q^2<25$ GeV$^{-2}$
are shown in Fig. 7. Here we presented
results for both $N_f=3$ and 4. As one observes the difference is quite
insignificant.

\begin{figure}[htbp]
  \begin{center}
    \epsfig{file=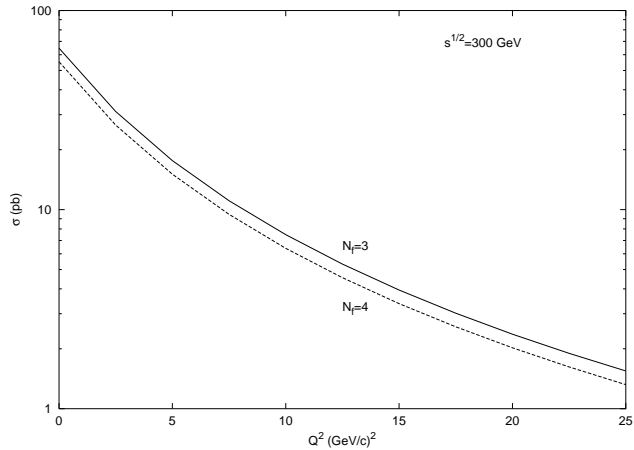,height=6cm} \bigskip
    \caption{Integrated cross-sections $\sigma$ for different number
  for flavors as a function of $Q^2$.}
\end{center}
\end{figure}

With the growth of energy both contributions increase, preserving their
shape in $q$. The increase is much more pronounced in the reggeizing
pomeron contribution, since in this case the pomeron occupies
the whole rapidity
range, whereas for the POO transition this range is shorter.

\section{Conclusions}
We have calculated the cross section of inclusive diffractive photo- and
leptoproduction of $\eta_c$ mesons, $\gamma^*p \to \eta_c +X$.
We have considered the `triple Regge' contribution which contains the
coupling POO of the pomeron to two odderons.
Inclusion of a second contribution where the pomeron directly couples to
two three-gluon states results in a significant rise
of the cross-section which grows with energy. However, in order to see the
structure of the QCD odderon state with $C=-1$ state one has to select
diffractive events with a large enough gap between the missing mass state
``X'' and the $\eta_c$. The total production rate is found to be of the
order 60 pb for photoproduction.

In our previous publication \cite{BBCV} we calculated the production
rate for the quasielastic reaction $\gamma^* +p\to \eta_c + p$ due to
odderon exchange. The photoproduction cross-section was reported
to be 27 pb, which would be smaller by a factor of $2$ compared to the
present case.
However, note that for the quasielastic process
considered in \cite{BBCV} one had to make an assumption of the
non-perturbative odderon-proton coupling. In \cite{BBCV} we used the coupling
proposed in \cite{Czyzewski:1997bv}, and we put the effective coupling
constant $\alpha_s$ equal to unity. However, in a recent analysis of the
$pp$ and $p\bar p$ elastic scattering data this coupling constant has been
estimated to be 0.3 \cite{Dosch}. With this value of the effective coupling
constant the cross-sections reported in \cite{BBCV} have to be reduced by a
factor 30, and, in fact, the cross section of the inclusive $\eta_c$
is much larger than the quasielastic one.

In the present calculation the very poorly known
odderon-proton coupling does not enter. Instead one has to know also the
non-perturbative pomeron-proton coupling, transformed into the value of
the product $\alpha_sR$ where $R$ is the
effective proton radius. This product can be found with a much
higher degree of reliability. In this study we have used the
fit to the experimental proton structure function in ~\cite{AB}.
Note that it gives   physically
reasonable values $\Delta=0.377$ and $R=0.59$ fm (for $N_f=4$),
which more or less agree with  estimates made by different methods.
Correspondingly, we feel that the cross-sections
found in this paper are much less affected by the uncertainty in
the nonperturbative coupling of the proton.

\section{Acknowledgements}
M.A.B. is deeply thankful to the University of Hamburg for
hospitality and financial support. This study was also partially
supported by the RFFI (Russia) grant 01-02-17137.
G.P.V. thanks the II. Inst. f\"ur Theor. Physik of the Hamburg
University for the warm hospitality.



\begin{thebibliography}{99}

\bibitem{Lukaszuk:1973nt}
L.~Lukaszuk and B.~Nicolescu,
Lett.\ Nuovo Cim.\ {\bf 8} (1973) 405.

\bibitem{Czyzewski:1997bv}
J.~Czyzewski, J.~Kwiecinski, L.~Motyka and M.~Sadzikowski,
Phys.\ Lett.\ {\bf B398} (1997) 400 [hep-ph/9611225]; erratum Phys. Lett
                {\bf B411} (1997) 402.

\bibitem{Bartels:2000yt}
J.~Bartels, L.~N.~Lipatov and G.~P.~Vacca,
Phys.\ Lett.\ {\bf B477} (2000) 178 [hep-ph/9912423].

\bibitem{Vacca:2000bk}
G.~P.~Vacca,
Phys.\ Lett.\ {\bf B489} (2000) 337 [hep-ph/0007067].

\bibitem{BBCV}
J.~Bartels, M.~A.~Braun, D.~Colferai and G.~P.~Vacca,
Eur.\ Phys.\ J.\ C {\bf 20} (2001) 323
[arXiv:hep-ph/0102221].


\bibitem{Heidelberg}
E.~R.~Berger, A.~Donnachie, H.~G.~Dosch, W.~Kilian, O.~Nachtmann and M.~Rueter,
Eur.\ Phys.\ J.\ {\bf C9} (1999) 491 [hep-ph/9901376].

A.~Schafer, L.~Mankiewicz and O.~Nachtmann,
UFTP-291-1992. 
In ``Hamburg 1991, Proceedings, Physics at HERA, vol. 1'' 243-251 and
Frankfurt Univ. - UFTP 92-291 (92,rec.Mar.) 8 p.

W.~Kilian and O.~Nachtmann,
Eur.\ Phys.\ J.\ {\bf C5} (1998) 317 [hep-ph/9712371].

M.~Rueter, H.~G.~Dosch and O.~Nachtmann,
Phys.\ Rev.\ D {\bf 59} (1999) 014018 [hep-ph/9806342].

\bibitem{BRV}
J.~Bartels, M.~G.~Ryskin and G.~P.~Vacca,
Eur.\ Phys.\ J.\ C {\bf 27} (2003) 101
[arXiv:hep-ph/0207173].

\bibitem{BE} J.Bartels and C.Ewerz, JHEP. {\bf 9}
             (1999) 26 [hep-ph/9908454]

\bibitem{BW} J.Bartels and M.Wuesthoff, Z.Phys {\bf C 66} (1995) 157.

\bibitem{BV}
M.~A.~Braun and G.~P.~Vacca,
Eur.\ Phys.\ J.\ C {\bf 6} (1999) 147
[arXiv:hep-ph/9711486].

\bibitem {AB} N.Armesto and M.A.Braun, Z.Phys {\bf C 76} (1997) 81.

\bibitem{Dosch}
H.~G.~Dosch, C.~Ewerz and V.~Schatz,
Eur.\ Phys.\ J.\ C {\bf 24} (2002) 561.
[arXiv:hep-ph/0201294].

\end{thebibliography}
\end{document}